# A step-by-step guide to generalized estimating equations using SPSS in the field of dentistry

Hoi-Jeong Lim
*Department of Statistics,*
*Department of Orthodontics, Chonnam National University School of*
*Dentistry, Dental Science Research Institute*
*Chonnam National University*
*Gwangju, South Korea*

Soo-Hyeon Park
*Department of Statistics,*
*Department of Orthodontics, Chonnam National University School of*
*Dentistry, Dental Science Research Institute*
*Chonnam National University*
*Gwangju, South Korea*

*Abstract*— The Generalized Estimating Equations (GEE) approach is a widely used statistical method for analyzing longitudinal data and clustered data in clinic studies. In dentistry, due to multiple outcomes obtained from one patient, the outcomes produced from individual patients are correlated with one another. This study focuses on the basic ideas of GEE and introduces the types of covariance matrix and working correlation matrix. The quasi-likelihood information criterion(QIC) and quasi-likelihood information criterion approximation(QICu) were used to select the best working matrix and the best fitting model for the correlated outcomes.

## I. Introduction

Statistical methods are determined by the scale of the outcome variable. When the outcome variable is continuous and we want to predict it using one or more explanatory variables, regression analysis is used. In cases where the outcome variable is categorical, such as death/survival or presence/absence of a disease, logistic regression analysis has been used to analyze each independent univariate outcome variable formed in this way. However, when the outcome variable is measured repeatedly on the same subjects (longitudinal data) or is derived from cluster sampling, there will be correlations between the repeated measures of the outcome variables or among the outcome variables derived from clustered data, which must be taken into account in statistical analysis. In the field of dentistry, when obtaining outcome values for multiple teeth within a single individual's mouth or studying the stability of multiple mini-screws implanted in a single individual's gums, there exists a correlation among the stability of the multiple mini-screws, which constitutes clustered data. Therefore, in such cases, instead of using logistic regression analysis, which cannot include the correlation between outcome variables, the Generalized Estimating Equations (GEE) analysis method should be applied, which assumes a correlation matrix and considers the correlation in the analysis process. The purpose of this study is to explore the procedures for performing Generalized Estimating Equations analysis in detail and to investigate how GEE analysis has been applied in dental research using SPSS.

## II. Basic concepts of GEE

The generalized estimating equations developed by Liang and Zeger(1986)[1] and Zeger and Liang(1986)[2] take into account the correlation between dependent variables. This method is an extension of the generalized linear model(GLM)[3], where each outcome is assumed to be independent. It assumes a working correlation matrix that represents the correlation between dependent variables, allowing for the estimation of parameters while considering the correlation among repeatedly measured outcome variables.[4]

In the case where $y_{ij}$, which has different times or locations for entity $i$ and $j$, is a categorical or continuous outcome variable, and $\chi = (\chi_{ji}, \cdots, \chi_{jip})'$ is a data structure consisting of a p×1 vector of categorical or continuous explanatory variables, the generalized estimating equation for estimating the coefficient β is as follows[1, 2].

$$\sum_{i=1}^{n} \frac{\partial \mu'}{\partial \beta} V_i^{-1}(Y_i - \mu_i(\beta)) = 0$$

Here, $\mu_i = (\mu_{i1}, \cdots, \mu_{it})'$ is the mean vector, and $Y_i = (y_{i1}, y_{i2}, \cdots, y_{it})$ is the response variable described above. $V_i$ is the estimator of the covariance matrix of $Y_i$, and it is written as follows.

$$V_i = \emptyset A_i^{1/2} R_i(\alpha) A_i^{1/2}$$

In the above formula, Ø refers to the overdispersion parameter, A refers to the diagonal matrix of variance functions, and R(α) refers to the pseudo-correlation matrix of Y.

### A. Covariance matrix

In GEE, there are two types of covariance matrices: the model-based covariance matrix and the robust covariance matrix. The robust covariance matrix is also referred to as the empirical or sandwich covariance matrix. The difference between these two covariance matrices is that the model-based covariance matrix provides consistent estimates of the model parameters and their variances only when the working correlation matrix is correctly specified. In contrast, the robust covariance matrix provides consistent estimates even when the working correlation matrix is not correctly specified, as long as the model for the mean µ of Y is correct. Researchers need to specify the working correlation matrix to fit the data, but errors can occur in selecting the appropriate working correlation matrix. Therefore, it is advisable to use the robust covariance matrix estimates, which are always consistent even if the working correlation matrix is not correctly specified[5]. The standard errors in GEE are obtained by taking the square root of each value of the robust covariance matrix. However, this estimator may be biased when the sample size is small or when the covariates have a skewed distribution, which can lead to bias (the deviation of the distribution center (mean) of the measurements or estimates from the true value[6])4].

## B. Working correlation matrix

The working correlation matrix represents the correlation between dependent variables obtained at different times from each subject. To estimate parameters, the researcher must specify a correlation matrix that fits the nature of the data. However, since the true correlation matrix is unknown, a fictitious correlation matrix corresponding to the table below is assumed, and this correlation matrix is treated as if it were the true correlation matrix for parameter estimation. When the working correlation matrix is not correctly specified, the relative efficiency of the parameter estimates obtained by the GEE method decreases. To address this issue, several scholars have proposed new criteria for selecting the working correlation matrix. Pan (2001)[7] proposed the QIC(Quasi-likelihood Information Criterion), Rotnitzky and Jewell (1990)[8] proposed the RJC(Rotnitzky and Jewell Criterion), Hin and Wang (2009)[9] proposed the CIC(Correlation Information Criterion), and Gosho et al. (2011)[10] proposed the DEW (Discrepancy between the covariance matrix Estimator and the specified Working correlation matrix). This study only addresses the QIC included in the SPSS results. The criterion for selecting the most appropriate working correlation matrix can be the lowest value of QIC, as explained in the next section. The working correlation matrix is classified into (1) Independent, (2) M-dependent, (3) Exchangeable, (4) Autoregressive (AR-1), (5) Unstructured, and (6) Fixed working correlation matrices, as shown in "Table 1" below. (1) assumes that the dependent variables obtained at different times from each subject are independent of each other, (2) assumes that correlation exists only up to the M-th observation, after which they are independent, (3) assumes that the degree of correlation between dependent variables obtained at different times from each subject is always constant at α, (4) indicates that the degree of correlation decreases over time, (5) assumes that there is no specific pattern and that the degree of correlation between all dependent variables is different for each, and (6) refers to the researcher specifying the working correlation matrix themselves[11].

Previously, independent working correlation matrices were primarily used in GEE analysis, but research on how to select the working correlation matrix has continued to this day. In many cases, GEE estimates assuming an independent working correlation matrix have been reported to have high efficiency. It has been noted that as the number of subjects increases, the efficiency in cases where the working correlation matrix is incorrectly specified (where a smaller variance indicates high efficiency and a larger variance indicates low efficiency; efficient estimators are those that are minimum variance unbiased estimators, meaning they have the smallest variance among unbiased estimators[12]) is reduced[13,14]. However, it has also been reported that using an independent working correlation matrix can lead to inefficient estimates when including time-varying covariates[15].

## C. QIC vs. QICu

In GEE, it is necessary to choose which working correlation matrix to use to explain the correlation between dependent variables and to determine which model fits best as the final model.

GLM is based on the maximum likelihood method[16], which estimates the parameters (θ) of a population by maximizing the likelihood function [L(θ)] [17]. On the other hand, the GEE method is based on the quasi-likelihood method, which is used when it is not possible to calculate the likelihood due to the absence of distribution assumptions for the observations[12, 18, 19]. In this case, the quasi-likelihood function can be used instead of the likelihood. The quasi-likelihood method estimates the parameters (θ) of a population by maximizing the quasi-likelihood function and does not make any assumptions about the distribution of the outcome variables. Therefore, statistics derived from likelihood theory cannot be directly applied to GEE. For example, AIC (Akaike's Information Criterion) is a widely used method for model selection in GLM, but it is not suitable for GEE. Consequently, Pan[7] modified the AIC method to create QIC and QICu, which are appropriate model selection methods for GEE analysis. The lowest QIC serves as a criterion for selecting the most suitable working correlation matrix, while the lowest QICu serves as a criterion for selecting the best-fitting model[20]. In the example used in this study, the QIC and QICu were employed to select the most suitable working correlation matrix, which was an unstructured working correlation matrix, and to build the final model that included age, location (maxilla/mandible), and implantation experience.

TABLE I. TYPES OF WORKING CORRELATION MATRIX

| (1) Independent working correlation matrix | (2) M-dependent working correlation matrix (M=2) | (3) Exchangeable working correlation matrix | (4) Autoregressive (AR-1) working correlation matrix | (5) Unstructured working correlation matrix |
|---|---|---|---|---|
| $R=\begin{pmatrix} 1 & 0 & 0 & 0 \\ 0 & 1 & 0 & 0 \\ 0 & 0 & 1 & 0 \\ 0 & 0 & 0 & 1 \end{pmatrix}$ | $R=\begin{pmatrix} 1 & \alpha_1 & \alpha_2 & 0 \\ \alpha_1 & 1 & \alpha_1 & \alpha_2 \\ \alpha_2 & \alpha_1 & 1 & \alpha_1 \\ 0 & \alpha_2 & \alpha_1 & 1 \end{pmatrix}$ | $R=\begin{pmatrix} 1 & \alpha & \alpha & \alpha \\ \alpha & 1 & \alpha & \alpha \\ \alpha & \alpha & 1 & \alpha \\ \alpha & \alpha & \alpha & 1 \end{pmatrix}$ | $R=\begin{pmatrix} 1 & \alpha & \alpha^2 & \alpha^3 \\ \alpha & 1 & \alpha & \alpha^2 \\ \alpha^2 & \alpha & 1 & \alpha \\ \alpha^3 & \alpha^2 & \alpha & 1 \end{pmatrix}$ | $R=\begin{pmatrix} 1 & \alpha_{21} & \alpha_{31} & \alpha_{41} \\ \alpha_{21} & 1 & \alpha_{32} & \alpha_{42} \\ \alpha_{31} & \alpha_{32} & 1 & \alpha_{43} \\ \alpha_{41} & \alpha_{42} & \alpha_{43} & 1 \end{pmatrix}$ |
| $Corr(y_{ij}, y_{i,j+s})$ $=\begin{cases} 1 & s=0 \\ 0 & s \neq 0 \end{cases}$ | $Corr(y_{ij}, y_{i,j+s})$ $=\begin{cases} 1 & s=0 \\ \alpha_s & s=1,2,\cdots,M \\ 0 & s>M \end{cases}$ | $Corr(y_{ij}, y_{i,j+s})$ $=\begin{cases} 1 & s=0 \\ \alpha & s \neq 0 \end{cases}$ | $Corr(y_{ij}, y_{i,j+s}) = \alpha^s$ $(s=0,1,2,\cdots,t_i-j)$ | $Corr(y_{ij}, y_{i,j+s})$ $=\begin{cases} 1 & s=0 \\ \alpha_{jk} & s \neq 0 \end{cases}$ |

III. INFORMATION ABOUT THE DATA USED IN DENTAL RESEARCH PAPER

In the past, headgear was used to provide orthodontic force during orthodontic treatment, but this method required the cooperation of the patient. As an alternative to address the disadvantages of headgear, mini screws were developed, which do not require patient cooperation and offer advantages such as ease of placement and low cost, and are primarily used to provide orthodontic force. However, due to the small size of mini screws, their stability during placement is lower, leading to an increase in studies investigating the factors that affect this stability. The following data is a hypothetical dataset obtained from a study investigating the factors influencing mini screw placement failure (LOOSENING) [21-23]. The example data for mini screws has been uploaded to the Graduate School of Dentistry at Chonnam National University, in the data room number 21, and the website address is provided below.

http://dent.jnu.ac.kr/user/indexSub.action?codyMenuSeq=6806&siteId=dent&menuUIType=top

ID = Patient identification number
AGE = Patient age
GENDER = Patient gender (Male=1, Female=0)
AREA1 = Mini screw implantation site (Maxilla=1, Mandible=0)
AREA2 = Specific mini screw implantation site (between the mandibular 1st and 2nd molars=1, between the maxillary 1st and 2nd molars=2, between the mandibular 1st molar and 2nd premolar=3, between the maxillary 1st molar and 2nd premolar=4, between the mandibular 1st and 2nd premolars=5, between the maxillary 1st and 2nd premolars=6, between the mandibular 1st premolar and canine=7, between the maxillary 1st premolar and canine=8, between mandibular canine and lateral incisor=9, between maxillary canine and lateral incisor=10, between mandibular central incisors=11, between maxillary central incisors=12)
LENGTH = Mini screw length (6mm, 7mm, 8mm, 10mm, 12mm)
DIAMETER = Mini screw diameter (1.6mm, 1.8mm)
NINSERT = Number of mini screw insertions by the doctor
DOCTOR_ID = Doctor identification number
LOOSENING = Mini screw insertion failure status (failure = 1, success = 0)
AGE1 = 1 if AGE>20, AGE1 = 0 if AGE≤20 LENGTH1 = 0 if LENGTH<8mm, LEN GTH1 = 1 if LENGTH≥8mm
NINSERT1 = 1 if NINSERT> 20, NIN SERT1 = 0 if NINSERT≤20

The outcome variable here is LOOSENING, and the explanatory variables are AGE, GENDER, AREA1, LENGTH, DIAMETER, and NINSERT.

Among these, the categorical variables (factors) are GENDER, AREA1, LENGTH, and DIAMETER, while the continuous variables (covariates) are AGE and NINSERT. For the sake of ease in interpretation, these continuous variables have been converted into categorical variables AGE1 and NINSERT1, and LENGTH has also been transformed into a binary variable LENGTH1 based on a threshold of 8mm.

Looking at the characteristics of this data, as shown in "Table 2", there were cases where one mini screw was implanted per patient, as well as cases with two, three, four, five, and six mini screws. Each patient forms a cluster with multiple outcomes (the stability of the mini screws). When examining whether the outcome variables within this cluster (here referred to as individuals) are highly correlated, "Table 3" shows that excluding the case where only one mini screw was implanted, the rate of all screws being successful (all success) was 59.6%, while the rate of all screws failing (all failure) was 3.8%. Additionally, cases that were almost successful or failed (skewed) accounted for 18.3%, and cases where success and failure were equally distributed also accounted for 18.3%. This indicates that if one mini screw is successful within an individual, there is a high probability that the other mini screws will also be successful, and if one fails, there is a high probability that the others will also fail. This shows that over 80% of the outcomes within individuals were highly correlated.

TABLE II. NUMBER OF MINISCREWS IMPLANTED FOR EACH PATIENT

| Number of miniscrews | Number of patients (%) |
|---|---|
| 1 | 31 (23.0) |
| 2 | 67 (49.6) |
| 3 | 17 (12.6) |
| 4 | 14 (10.4) |
| 5 | 3 (2.2) |
| 6 | 3 (2.2) |
| Total | 135 (100) |

TABLE III. INDIVIDUAL PATIENT'S MINISCREW STABILITY DISTRIBUTION EXCEPT PATIENTS IMPLANTED ONE MINISCREW

| Distribution | Number of patients (%) |
|---|---|
| All success | 62 (59.6) |
| All failure | 4 (3.8) |
| Skewed | 19 (18.3) |
| Equally distributed | 19 (18.3) |
| Total | 104 (100) |

IV. DESCRIPTION OF SPSS PROCEDURES AND RESULTS

*A. SPSS procedure for univariate GEE analysis*

The following is a description of the SPSS procedure for GEE univariate analysis in "Fig. 1".

*1) The data used here is as follows.*

*2) Click on Analyze > Generalized Linear Model > Generalized Estimating Equations.*

*3) Move the variable ID to the subject variable and AREA2 to the within-subject variable, then select 'Unstructured' in the working structure of the correlation matrix, and click on the model type in the menu.*

*4) In the custom settings, select Distribution: Binomial, Link function: Logit, and then click on the response in the menu. Alternatively, you can also select Binary response or Binary logistic for event/trial data and then click on the response in the menu to obtain the same result.*

*5) Move the dependent variable LOOSENING, click on the reference category, and select the first (lowest value). After clicking continue, click on predictors in the menu.*

*6) Move the factor AREA1, click on options, select descending order in the category order for the factor, and click continue.*

*7) Select model from the menu, and use the arrow to move AREA1 into the model under factors and covariates.*

*8) Select statistics from the menu, and choose the statistics as shown below in the print options.*

To explain the above process in more detail, when the response variable is continuous (normally distributed response) as shown in "Fig. 1(4)", the distribution is chosen as 'normal' and the link function is 'identity (I)'. If the response variable consists of 0/1, the distribution is chosen as 'binomial test' and the link function is 'logit log-linear analysis'. Additionally, if the response variable is the count of occurrences per unit time), the distribution is chosen as 'Poisson distribution (I)' and the link function is 'log'.

"Fig. 1(5)" defines the reference category for the binary response variable (LOOSENING=0 or 1). The reference category refers to the category that serves as a baseline for comparison among categories. Here, the first (lowest value) is LOOSENING=0 (success), which is the reference category, while the last (highest value) is LOOSENING=1 (failure), which becomes the reference category.

In "Fig. 1(6)", clicking on the options under the factor opens a new window. There, the category order for the factor (AREA1=0 or 1) defines the reference category of the factor, and the latter part of the order becomes the reference category. In the case of ascending order, since 1 is the latter part of the order between 0 and 1, AREA1=1 (maxilla) becomes the reference category. In the case of descending order, since 0 is the latter part of the order between 1 and 0, AREA1=0 (mandible) becomes the reference category.

"Fig. 2" shows the results of univariate analysis when assuming an unstructured covariance matrix for the AREA1 variable. In "Fig. 2(1)", when the reference variable of the dependent variable is LOOSENING=0, OR=Exp(B)=0.54, and the 95% confidence interval for OR is (0.30, 1.00)*. Since AREA1=1 (maxilla) is the reference variable, this means that the risk of failure in the maxilla is 0.54 times lower compared to the mandible. "Fig. 2(2)" changes the reference variable of the dependent variable from LOOSENING=0 to LOOSENING=1, indicating that the risk of success in the maxilla is 1.84 (=1/0.54) times higher than that in the mandible.

"Fig. 3" examines the changes in the OR values according to the reference category of the dependent variable in "Fig. 1(5)" and the reference category of the factors in "Fig. 1(6)" when using an independent virtual correlation matrix.

*a) In the type of dependent variable in "Fig. 1(5)" (only applicable to the binomial distribution), the reference category is the first (the lowest value). If descending order is selected for the categories of the factor in "Fig. 1(6)", then LOOSENING=0 becomes the reference category for the dependent variable, and AREA=0 becomes the reference category for the factor. In this case, it is*

$$OR = \frac{ad}{bc} = \frac{42 \times 52}{184 \times 27} = 0.44$$

*b) In the dependent variable type of "Fig. 1(5)" (only applicable to the binomial distribution), the reference category is the first (the lowest value). If you select ascending order for the categories of the factor in "Fig. 1(6)", LOOSENING=0 becomes the reference category for the dependent variable, and AREA=1 becomes the reference category for the factor. In this case, it is*

$$OR = \frac{ad}{bc} = \frac{27 \times 184}{52 \times 42} = 2.27$$

*c) In the type of dependent variable in "Fig. 1(5)" (only applicable to the binomial distribution), the reference category is the last (highest value). If you select descending order for the categories of the factor in "Fig. 1(6)", LOOSENING=1 becomes the reference category for the dependent variable, and AREA=0 becomes the reference category for the factor. In this case, it is*

$$OR = \frac{ad}{bc} = \frac{184 \times 27}{42 \times 52} = 2.27$$

*d) In the dependent variable type of "Fig. 1(5)" (only applicable to the binomial distribution), the reference category is the last (highest value). In "Fig. 1(6)", if ascending order is selected for the categories of the factors, LOOSENING=1 becomes the reference category for the dependent variable, and AREA=1 becomes the reference category for the factor. In this case, it is*

$$OR = \frac{ad}{bc} = \frac{52 \times 42}{27 \times 184} = 0.44$$

Odds are defined as the ratio of the probability of failure to the probability of success (how many times). The concept of odds is not commonly used in our country, making it difficult to understand. The meaning of odds in the maxilla is odds = 42/184 = 1/4.38 = 0.2283, which indicates that the probability of failure in the maxilla is 0.23 times the probability of success. In other words, for every 4.38 successes, we can expect 1 failure. Additionally, the meaning of odds in the mandible is odds = 27/52 = 1/1.92 = 0.52, which indicates that the probability of failure in the mandible is 0.52 times the probability of success. This means that for every 1.92 successes, we can expect 1 failure. The odds ratio is obtained by dividing these two odds (OR = 0.23 / 0.52 = 0.44), which means that the risk of failure in the maxilla is 0.44 times higher than in the mandible, indicating that the risk of success in the maxilla is 2.27 times higher than in the mandible.

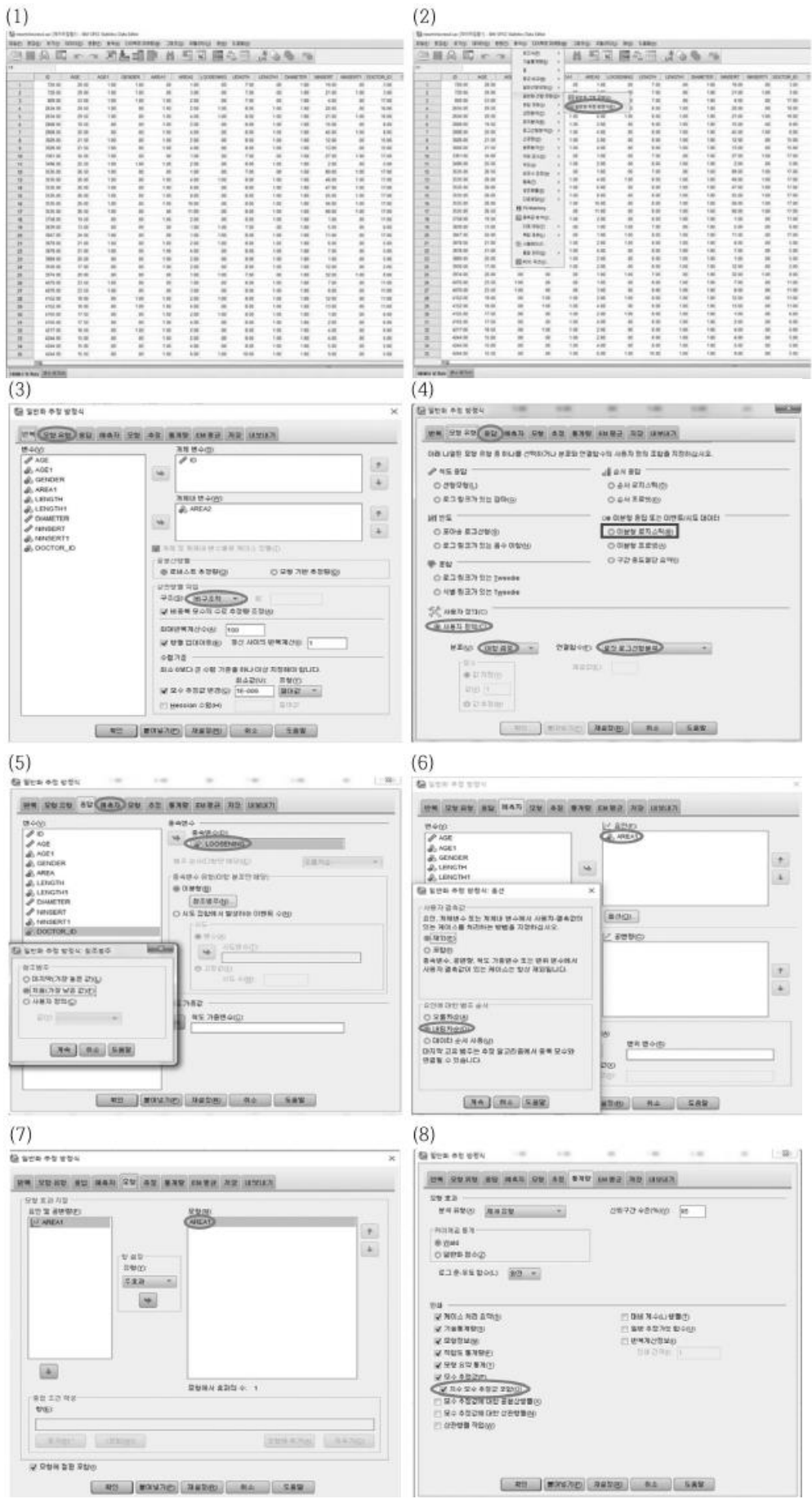

Fig. 1. Univariable analysis of Generalized Estimating Equations

(1) SPSS outcome with miniscrew LOOSENING = 0 as reference category

| 모수 | B | 표준오차 | 95% Wald 신뢰구간 하한 | 95% Wald 신뢰구간 상한 | Wald 카이제곱 | 자유도 | 유의확률 | Exp(B) | Exp(B)에 대한 95% Wald 신뢰구간 하한 | Exp(B)에 대한 95% Wald 신뢰구간 상한 |
|---|---|---|---|---|---|---|---|---|---|---|
| (절편) | -.756 | .2677 | -1.280 | -.231 | 7.970 | 1 | .005 | .470 | .278 | .794 |
| [AREA1=1.00] | -.609 | .3113 | -1.219 | .001 | 3.828 | 1 | .050 | .544 | .295 | 1.001 |
| [AREA1=.00] | 0a | . | . | . | . | . | . | 1 | . | . |
| (척도) | 1 | | | | | | | | | |

종속변수: LOOSENING
모형: (절편), AREA1
a. 중복된 모수이므로 0으로 설정됩니다.

(2) SPSS outcome with miniscrew LOOSENING = 1 as reference category

| 모수 | B | 표준오차 | 95% Wald 신뢰구간 하한 | 95% Wald 신뢰구간 상한 | Wald 카이제곱 | 자유도 | 유의확률 | Exp(B) | Exp(B)에 대한 95% Wald 신뢰구간 하한 | Exp(B)에 대한 95% Wald 신뢰구간 상한 |
|---|---|---|---|---|---|---|---|---|---|---|
| (절편) | .756 | .2677 | .231 | 1.280 | 7.970 | 1 | .005 | 2.129 | 1.260 | 3.598 |
| [AREA1=1.00] | .609 | .3113 | -.001 | 1.219 | 3.828 | 1 | .050 | 1.839 | .999 | 3.384 |
| [AREA1=.00] | 0a | . | . | . | . | . | . | 1 | . | . |
| (척도) | 1 | | | | | | | | | |

종속변수: LOOSENING
모형: (절편), AREA1
a. 중복된 모수이므로 0으로 설정됩니다.

Fig. 2. SPSS outcome after univariable analysis for each reference category using unstructured working correlation matrix

(1) OR=0.44

| LOOSENING \ AREA | 1 | 0 |
|---|---|---|
| 1 | 42 | 184 |
| 0 | 27 | 52 |

(2) OR=2.27

| LOOSENING \ AREA | 1 | 0 |
|---|---|---|
| 0 | 27 | 52 |
| 1 | 42 | 184 |

(3) OR=2.27

| LOOSENING \ AREA | 0 | 1 |
|---|---|---|
| 1 | 184 | 42 |
| 0 | 52 | 27 |

(4) OR=0.44

| LOOSENING \ AREA | 0 | 1 |
|---|---|---|
| 0 | 52 | 27 |
| 1 | 184 | 42 |

Fig. 3. 2x2 contingency tables by various reference categories

(1) SPSS outcome with miniscrew LOOSENING = 0 as reference category

| 모수 | B | 표준오차 | 95% Wald 신뢰구간 하한 | 95% Wald 신뢰구간 상한 | Wald 카이제곱 | 자유도 | 유의확률 | Exp(B) | Exp(B)에 대한 95% Wald 신뢰구간 하한 | Exp(B)에 대한 95% Wald 신뢰구간 상한 |
|---|---|---|---|---|---|---|---|---|---|---|
| (절편) | -.655 | .3321 | -1.306 | -.004 | 3.894 | 1 | .048 | .519 | .271 | .996 |
| [AREA1=1.00] | -.822 | .3800 | -1.567 | -.077 | 4.678 | 1 | .031 | .440 | .209 | .926 |
| [AREA1=.00] | 0a | . | . | . | . | . | . | 1 | . | . |
| (척도) | 1 | | | | | | | | | |

종속변수: LOOSENING
모형: (절편), AREA1
a. 중복된 모수이므로 0으로 설정됩니다.

(2) SPSS outcome with miniscrew LOOSENING = 1 as reference category

| 모수 | B | 표준오차 | 95% Wald 신뢰구간 하한 | 95% Wald 신뢰구간 상한 | Wald 카이제곱 | 자유도 | 유의확률 | Exp(B) | Exp(B)에 대한 95% Wald 신뢰구간 하한 | Exp(B)에 대한 95% Wald 신뢰구간 상한 |
|---|---|---|---|---|---|---|---|---|---|---|
| (절편) | .655 | .3321 | .004 | 1.306 | 3.894 | 1 | .048 | 1.926 | 1.004 | 3.693 |
| [AREA1=1.00] | .822 | .3800 | .077 | 1.567 | 4.678 | 1 | .031 | 2.275 | 1.080 | 4.791 |
| [AREA1=.00] | 0a | . | . | . | . | . | . | 1 | . | . |
| (척도) | 1 | | | | | | | | | |

종속변수: LOOSENING
모형: (절편), AREA1
a. 중복된 모수이므로 0으로 설정됩니다.

Fig. 4. SPSS outcome after univariable analysis for each reference category using independent working correlation matrix

Here, we found that the OR values change depending on the reference category of the dependent variable and the reference category of the factors, which makes the interpretation of the results easier to understand. "Fig. 4" shows the results obtained using SPSS for univariate analysis of the AREA1 variable, changing the reference category from 0 to 1 under the assumption of an independent virtual correlation matrix. We observed that the OR values in "Fig. 3" and "Fig. 4" are 0.44 and 2.27, respectively, which are consistent.

### B. SPSS Procedure for Multivariable GEE Analysis

Below is an explanation of the SPSS procedure for GEE multivariate analysis in "Fig. 5".

*1) Move AGE1, AREA1, and NINSERT1 to the factors, click on the options, and select descending order in the category order for the factors.*

*2) Select the model from the top menu and move AGE1, AREA1, and NINSERT1 to the model under factors and covariates.*

*3) Press confirm.*

"Fig. 6" shows the results of a multivariate analysis conducted using SPSS, and a detailed interpretation of the results can be found below "Table 5".

### C. Finding the best-fitting final model

*1) To select the most suitable virtual correlation matrix, "Table 4" is :*

  *a) STEP 1:* For each covariate, AGE1, GENDER, AREA1, LENGTH1, DIAMETER, and NINSERT1, the QIC values were calculated for all types of virtual correlation matrices. When there was an unstructured virtual correlation matrix for AREA1, it had the lowest QIC value of QIC=323.410 among the models that included one covariate, compared to when other virtual correlation matrices were used.

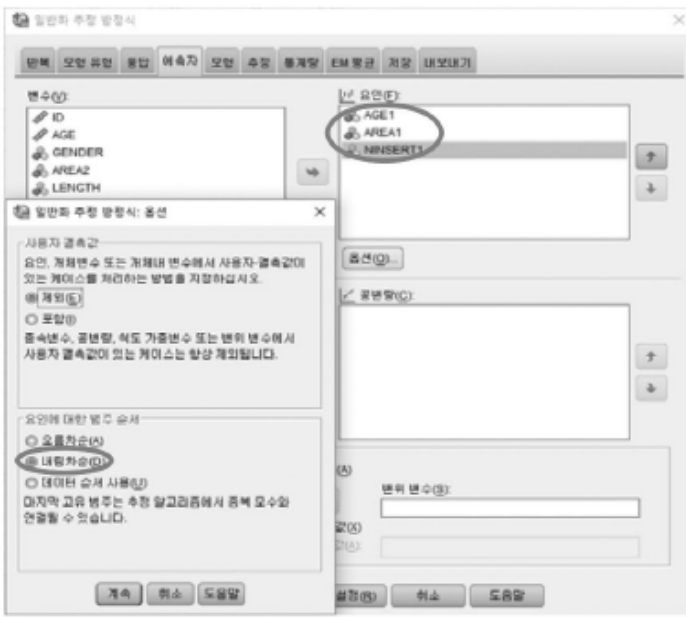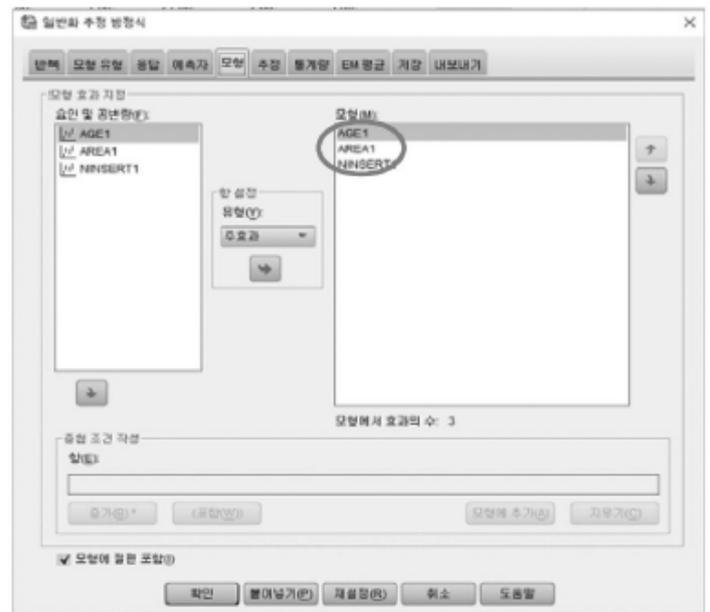

Fig. 5. Multivariable analysis of Generalized Estimating Equations

Fig. 6. SPSS outcome after multivariable analysis of Generalized Estimating Equations using unstructured working correlation matrix

*b) STEP 2*: In a model that includes two covariates, we calculate the QIC values for each of the corresponding virtual correlation matrices. The models that include two covariates are composed of (AGE1, GENDER), (AGE1, AREA1), (AGE1, LENGTH1), …, (DIAMETER, NINSERT1). As a result of calculating the QIC values for each virtual correlation matrix in the models that include two covariates, the model that includes AREA1 and NINSERT1 had a non-structural virtual correlation matrix and a QIC of 318.344, which is lower than the QIC value of the model that includes only the single covariate AREA1.

*c) STEP 3:* In a model that includes three covariates, we calculate the QIC values for each of the corresponding correlation matrices. A model that includes three covariates consists of combinations such as (AGE1, GENDER, AREA1), (AGE1, GENDER, LENGTH1), (AGE1, GENDER, DIAMETER), ..., (LENGTH1, DIAMETER, NINSERT1). The model with the three covariates AREA1, NINSERT1, and AGE1 had a lower QIC value (=315.080) compared to models that included only one covariate and those that included two covariates.

*d) STEP 4:* We calculate the QIC values for models that include four covariates, five covariates, and six covariates, respectively. As a result, the model with three covariates: AREA1, NINSERT1, and AGE1 has the lowest QIC value (=315.080), indicating that the unstructured covariance matrix is the most suitable.

*2) To select the most suitable model:*

*a) STEP 1:* Using the selected unstructured covariance matrix from above, find the lowest QICu value in a model with one covariate. As a result, the model including AREA1 was selected (QICu = 323.109).

*b) STEP 2:* Using the unstructured covariance matrix, find the model with the lowest QICu value for the model with two covariates. As a result, the model including AREA1 and NINSERT1 was selected (QICu = 318.215).

*c) STEP 3:* Using the unstructured virtual correlation matrix, find the model with the lowest QICu value among the models with three covariates. As a result, the model including AREA1, NINSERT1, and AGE1 was selected (QICu = 315.004).

*d) STEP 4:* Using the unstructured covariance matrix, find the lowest QICu value in models with four, five, and six covariates. As a result, the model with three covariates, AREA1, NINSERT1, and AGE1, had the lowest QICu value (=315.004) and was selected as the final model.

TABLE IV.    QICU FOR MODEL SELECTION UNDER BINOMIAL DISTRIBUTION AND LOGIT LINK

| Correlation | Variable | p* | QIC† | QIC$_u$‡ |
|---|---|---|---|---|
| Independent | AREA1, NINSERT1, AGE1 | 4 | 316.912 | 314.466 |
| Exchangeable | AREA1, NINSERT1, AGE1 | 4 | 316.651 | 314.499 |
| M-dependent (M=2) | AREA1, NINSERT1, AGE1 | 4 | 315.659 | 314.587 |
| Unstructured | AREA1, NINSERT1, AGE1 | 4 | **315.080** | 315.004 |
| Unstructured | AGE1 | 2 | 327.757 | 326.870 |
| Unstructured | GENDER | 2 | 327.071 | 327.022 |
| Unstructured | AREA1 | 2 | **323.410** | **323.109** |
| Unstructured | LENGTH1 | 2 | 328.520 | 328.281 |
| Unstructured | DIAMETER | 2 | 329.290 | 328.954 |
| Unstructured | NINSERT1 | 2 | 324.595 | 324.644 |
| Unstructured | AREA1, LENGTH1 | 3 | 324.959 | 324.984 |
| Unstructured | AREA1, DIAMETER | 3 | 320.545 | 320.029 |
| Unstructured | AREA1, NINSERT1 | 3 | **318.344** | **318.215** |
| Unstructured | AREA1, NINSERT1, AGE1 | 4 | **315.080** | **315.004** |
| Unstructured | AREA1, NINSERT1, GENDER | 4 | 317.469 | 317.742 |
| Unstructured | AREA1, NINSERT1, LENGTH1 | 4 | 319.658 | 320.018 |
| Unstructured | AREA1, NINSERT1, DIAMETER | 4 | 317.933 | 317.899 |
| Unstructured | AREA1, NINSERT1, AGE1, GENDER | 5 | 315.315 | 315.568 |
| Unstructured | AREA1, NINSERT1, AGE1, DIAMETER | 5 | 315.253 | **315.430** |

Note. Numbers in boldface indicate smallest QIC$_u$ value
*p, Number of parameters; †QIC, Quasi-likelihood under the independence model criterion; ‡QIC$_u$≈QIC when the GEE model is correctly specified.

TABLE V. THE CRUDE ODDS RATIO, ADJUSTED ODDS RATIO, AND THEIR 95 PERCENT CONFIDENCE INTERVALS (CIS) FOR THE CONTRIBUTING FACTORS FOR THE STABILITY THROUGH GENERALIZED ESTIMATING EQUATIONS

| Variables | Univariable cOR (95% CI) | Multivariable aOR (95% CI) |
|---|---|---|
| Gender | | |
|   Female | 1.00 | |
|   Male | 1.50 (0.86, 2.63) | |
| Age(years) | | |
|   ≤20 | 1.00 | 1.00 |
|   >20 | 0.68 (0.38, 1.21) | 0.58 (0.33, 1.02) |
| Jaw | | |
|   Mandible | 1.00 | 1.00 |
|   Maxilla | 0.54 (0.30, 1.00)* | 0.43 (0.23, 0.79)* |
| Diameter of miniscrews(mm) | | |
|   1.6 | 1.00 | |
|   1.8 | 1.93 (0.53, 7.10) | |
| Length of miniscrews(mm) | | |
|   <8 | 1.00 | |
|   ≥8 | 0.78 (0.42, 1.44) | |
| Number of insertions | | |
|   ≤20 | 1.00 | 1.00 |
|   >20 | 0.50 (0.28, 0.91)* | †0.48 (0.26, 0.89)* |

cOR: crude odds ratio, aOR: adjusted odds ratio,
†aOR=0.48: Age and jaw were adjusted using GEE for loosening variable.

### D. Results and Interpretation of Univariate and Multivariate GEE Analysis

To examine the results of the multivariate analysis in "Table 5", let's look at the adjusted odds ratios (OR). Patients over the age of 20 had 0.58 times less failure during implantation compared to those aged 20 and under, but this difference was not statistically significant. The maxilla showed a statistically significant 0.43 times less failure compared to the mandible, and those with more than 20 implantation experiences had 0.48 times less failure compared to those with no experience. In other words, the mandible and having no implantation experience based on the 20-case criterion were associated with a significantly higher failure rate in mini-screw implantation. The reason age was included in the final model is that it was close to 1 in the 95% confidence interval, indicating a lack of significance; however, it is believed that if the sample size were slightly larger, it might show significance. If the null hypothesis states that implantation experience is not related to mini-screw implantation failure, then even after controlling for the differences in failure rates due to age and the maxilla versus mandible, it can be concluded that having more than 20 implantation experiences significantly leads to the success of mini-screw implantation.

## V. CONCLUSION

GEE (Generalized Estimating Equations) analysis is an essential statistical method in dental research that involves multiple outcome observations for each patient. However, since the GEE analysis method falls under advanced statistical techniques that may not be easily understood by dental professionals who do not specialize in statistics, this work focuses only on the necessary procedures and theories to make it as accessible as possible for researchers. The explanations of the procedures for conducting GEE analysis using SPSS in various fields of dentistry aim to inform researchers who wish to analyze clustered data with multiple outcomes per patient, and to assist them in performing statistical analyses independently to obtain and interpret research results.